\documentclass[final,5p,times,twocolumn,authoryear]{elsarticle}

\usepackage{amssymb}
\usepackage{lipsum}
\usepackage[colorlinks = true, linkcolor = blue, urlcolor  = blue, citecolor = blue]{hyperref}
\usepackage{tabularx}

\newcommand{\pinocchio}{\texttt{PINOCCHIO}}
\newcommand{\CU}{\texttt{CU}}
\newcommand{\CUs}{\texttt{CUs}}

\journal{Astronomy $\&$ Computing}

\begin{document}

\begin{frontmatter}



\title{Accelerating cosmological simulations on GPUs: a portable approach using OpenMP}


\author[U_TS,OATS,ICSC,IFPU]{M. D. Lepinzan}
                  
\author[OABR,ICSC]{G. Lacopo}
            
\author[OATS,ICSC]{D. Goz}

\author[OATS,ICSC]{G. Taffoni}

\author[U_TS,OATS,ICSC,IFPU]{P. Monaco}

\author[PAWSEY]{P. J. Elahi}

\author[PAWSEY]{U. Varetto}

\author[PAWSEY]{M. Cytowski}

\affiliation[U_TS]{organization={Dipartimento di Fisica, Sezione di Astronomia, Università di Trieste},
            addressline={via Tiepolo 11}, 
            city={Trieste},
            postcode={34143}, 
            state={TS},
            country={Italy}}
\affiliation[OATS]{organization={INAF-Osservatorio Astronomico di Trieste},
            addressline={via Tiepolo 11}, 
            city={Trieste},
            postcode={34143}, 
            state={TS},
            country={Italy}}
\affiliation[OABR]{organization={INAF-Osservatorio astronomico di Brera},
            addressline={Via E. Bianchi, 46}, 
            city={Merate},
            postcode={23807}, 
            state={LC},
            country={Italy}}
\affiliation[ICSC]{organization={ ICSC -- Centro Nazionale di Ricerca in High Performance Computing, Big Data e Quantum Computing},
            addressline={via Magnanelli 2}, 
            city={Bologna},
            postcode={40033}, 
            state={TS},
            country={Italy}}
\affiliation[IFPU]{organization={IFPU -- Institute for Fundamental Physics of the Universe},
            addressline={via Beirut 2}, 
            city={Trieste},
            postcode={34151}, 
            state={TS},
            country={Italy}}
\affiliation[PAWSEY]{organization={Pawsey Supercomputing Centre}, 
            addressline={1 Bryce Avenue},
            city={Kensington WA},
            postcode={6151},
            country={Australia}}

\begin{abstract}
In this work we present the porting to Graphics Processing Units (GPUs, using OpenMP target directives) and optimization of a key module within the cosmological {\pinocchio} code, a Lagrangian Perturbation Theory (LPT)-based framework widely used for generating dark matter (DM) halo catalogs.

Our optimization focuses on a specific segment of the code responsible for calculating the collapse time of each particle involved in the simulation. Due to the embarrassingly parallel nature of this computation, it represents an ideal candidate for GPU offloading. 

As part of the porting process, we developed fully GPU-native implementations of both cubic spline and bilinear interpolation routines, required for evaluating collapse times. Since GNU Scientific Library (GSL) does not support GPU offloading, these custom implementations run entirely on the GPU and achieve residuals of only $\sim0.003\%$ when compared to the CPU-based implementation of GSL. 

Comparative benchmarking on the LEONARDO (NVIDIA-based) and SETONIX (AMD-based) supercomputers reveals notable portability and performance, with speedups of 4$\times$ and up to 8$\times$, respectively. While collapse time calculation is not a primary bottleneck in the overall workflow, the acceleration reduces full production runs by $\sim 100$ seconds each leading to a cumulative saving of $\sim 160000$ Standard-h ($\sim28$ hours wall time) across thousands of simulations.

Roofline analysis confirms that our GPU porting achieves over 80\% of the theoretical FP64 peak performance, confirming efficient compute-bound execution.

This work demonstrates that OpenMP directives offer a portable, effective strategy for accelerating large-scale cosmological simulations on heterogeneous hardware.
\end{abstract}




\end{frontmatter}




\section{Introduction}
The growing scale of simulations in computational cosmology demands increasingly efficient execution to align with the pace of advancements in scientific discovery. As High Performance Computing (HPC) platforms continue to evolve, leveraging their full potential, especially through the acceleration capabilities of Graphics Processing Units (GPUs), has become mandatory~\citep[see e.g.][]{space2, 10.1117/12.3020361}.

Several recent works have explored GPU acceleration and performance portability strategies for scientific computing applications, including astrophysical and cosmological codes. These efforts span approaches based on CUDA and MPI hybridization such as~\cite{Cesare_2023}, OpenMP offloading to heterogeneous architectures such as~\cite{Salvadore2024}, and broader studies of performance portability across diverse systems such as~\cite{Demeure_2023},~\cite{Malenza_2025} and~\cite{Bourramouss_2025}, reflecting a growing community emphasis on performance portability in modern HPC software development.

Within this landscape, the present work focuses on optimizing the fast cosmological simulation code {\pinocchio}~\citep{Monaco:2001jg, Monaco:2013qta}, a Lagrangian Perturbation Theory (LPT)-based framework widely used for generating DM halo catalogs. {\pinocchio} plays a crucial role in the generation of mock catalogs~\citep{Euclid:2025lfa} for large-scale surveys such as Euclid~\citep{Euclid:2024yrr}, where speed and scalability are critical to produce large number of realizations required for robust statistical analyses such as covariance matrices estimation and calibration~\citep{Euclid:2021api,Euclid:2022txd, Salvalaggio:2024vmx}.
 
To enhance both performance and portability, we offload a specific segment of the code to GPUs using OpenMP target directives. While not the primary computational bottleneck, this part of the code is embarrassingly parallel, making it an ideal candidate for GPU offloading. Since the original CPU-based implementation relies on the GNU Scientific Library (GSL)~\footnote{\url{https://www.gnu.org/software/gsl/doc/html/interp.html\#}} for interpolation routines, which does not support GPU offloading, we developed custom GPU-native interpolation routines to enable full device execution. This approach enables GPU acceleration ensuring portability across architectures, in the sense commonly adopted in the performance-portability literature~\citep{PENNYCOOK2019947}, including NVIDIA and AMD platforms, and long-term maintainability.

OpenMP~\citep{OpenMP}, with GPU support introduced in version 4.0, provides a directive-based programming model that enables the offloading of computations to GPUs while maintaining cross-platform portability and preserving code readability. Moreover, it avoids the need to introduce C++ dependencies typically required for CUDA~\citep{farber2011cuda} and HIP \footnote{\url{https://rocm.docs.amd.com/projects/HIP/en/docs-develop/what_is_hip.html}} models. This approach allows for gradual integration into existing CPU-based code, making it especially suitable for modernizing legacy scientific applications. While the directive-based model facilitates integration within the existing code, achieving optimal performance, particularly in the presence of complex control flow, often requires careful adaptation to GPU-friendly execution patterns.
 
Through comparative benchmarking on two major supercomputing platforms, LEONARDO (NVIDIA-based) and SETONIX (AMD-based), we show both the portability and significant performance gains of our approach, achieving speedup factors of at least 4$\times$ on the NVIDIA and up to 8$\times$ on the AMD platforms in single-node tests, the latter benefiting from its higher peak floating-point performance. This results underscore the potential of OpenMP-based GPU offloading to accelerate scientific simulations effectively across diverse hardware accelerators.

This work is organized as follows. Section~\ref{Computing_platforms} provides an overview of the key characteristics of the computing platforms used in this study. Section~\ref{Pinocchio_GPU_PMT} introduces the main components of the {\pinocchio} code, highlighting the embarrassingly parallel nature of the target segment. Section~\ref{Methodology} describes the methodology adopted for GPU porting using OpenMP, including the implementation of custom GPU-native interpolation routines required to replace the non-portable GSL-based functions. We also detail the validation of these GPU routines against their CPU counterparts, as well as the resource configurations adopted for both single-node tests and large-scale production runs. The results of our implementation are presented and discussed in Section~\ref{Results}, including the validation and performance assessment of the custom GPU interpolation routines using a standalone toy code, followed by single-node benchmarks within full {\pinocchio} runs on both NVIDIA and AMD platforms. The impact of GPU offloading in full-scale production runs is evaluated only on the NVIDIA platform, as large-scale resources were not available on the AMD system. Finally, conclusions are given in Section~\ref{Conclusion}.

\section{Computing platforms}
\label{Computing_platforms}

To assess the portability and performance of our GPU porting strategy, we conducted benchmarks on two state-of-the-art supercomputing platforms, described in detail below. These systems, LEONARDO and SETONIX, represent distinct GPU architectures (NVIDIA and AMD, respectively), offering an ideal environment to evaluate cross-platform compatibility and scalability.

\subsection{NVIDIA cluster}
Leonardo Booster~\citep{Turisini_2023}, part of the CINECA \footnote{\url{https://www.hpc.cineca.it/systems/hardware/leonardo/}} Italian national HPC facility, is ranked as $6^{th}$ in the November 2024 HPCG500 list and $52^{th}$ in the November 2024 Green500. The platform consists of 3456, 32-cores, Intel Xeon Platinum 8358 CPU nodes equipped with 4 NVIDIA Tesla Ampere 100 GPUs. Each CPU node has 512 GB of DDR4 memory, while each GPU provides 64 GB of HBM2 memory. The CPU and the GPU are interfaced by a PCIe Gen4 interconnection. 

The {\pinocchio} code was compiled on this system using the NVC/NVC++ compilers v24.3 and OpenMPI library v5.1 was employed.

\subsection{AMD cluster}
Setonix-GPU, part of the HPC facilities at the Pawsey Supercomputing Centre \footnote{\url{https://pawsey.org.au/systems/setonix/}} in Perth, Western Australia, is ranked as $45^{th}$ in the November 2024 HPCG500 list and $20^{th}$ in the November 2024 Green500 list.
The partition comprises 154 compute nodes with 256 GB memory each, equipped with 1 $\times$ AMD optimized 3rd Gen EPYC ``Trento" 64 cores and 8 GCDs (from 4x ``AMD MI250X" cards, each card with 2 GCDs), 128 GB HBM2e. CPU-GPU and GPU-GPU interconnections within each node node are guaranteed by the InfinityFabric technology. 

The {\pinocchio} code was compiled on this system with the amdclang-18 compiler and MPICH-8.1.31 implementation was employed.

\subsection{Architectural Comparison and Performance Implications}
\label{GPU_architectures}

To better understand the performance differences observed in our benchmarks, we provide a detailed analysis of the GPU architectures and their interconnection topologies on both platforms.

The NVIDIA A100 GPU is based on the Ampere architecture, featuring:
\begin{itemize}
    \item \textbf{Compute Units}: 108 Streaming Multiprocessors (SMs) with 6912 FP64 CUDA cores
    \item \textbf{FP64 Performance}: 9.7 TFLOPS peak theoretical performance
    \item \textbf{Memory Subsystem}: 64 GB HBM2 with 1555 GB/s bandwidth per GPU
    \item \textbf{Cache Hierarchy}: 40 MB L2 cache, 192 KB L1 cache per SM
    \item \textbf{Tensor Cores}: 432 third-generation Tensor Cores (not utilized in our application)
\end{itemize}

The AMD MI250X represents a unique multi-chip module (MCM) design (6nm process), where each card contains two Graphics Compute Dies (GCDs):
\begin{itemize}
    \item \textbf{Compute Units}: Each GCD has 110 Compute Units (CUs) with 7040 stream processors
    \item \textbf{FP64 Performance}: 47.9 TFLOPS peak theoretical performance per GCD (95.8 TFLOPS per card)
    \item \textbf{Memory Subsystem}: 64 GB HBM2e per GCD (128 GB per card) with 3276 GB/s bandwidth per GCD
    \item \textbf{Cache Hierarchy}: 8 MB L2 cache per GCD, 16 KB L1 cache per CU
    \item \textbf{Matrix Cores}: 440 Matrix Core Engines per GCD (not utilized in our application)
\end{itemize}

The FP64 peak performance differs by approximately 5$\times$ between AMD GCDs (47.9 TFLOPS) and NVIDIA A100s (9.7 TFLOPS). Such a disparity is expected to influence the execution of compute-bound kernels, with its impact examined later in Section~\ref{Results}. 

\subsubsection{CPU-GPU Interconnection Topology}

The LEONARDO nodes employ a traditional PCIe-based topology:

\begin{verbatim}
CPU Socket 0 (16 cores) <-PCIe Gen4-> GPU 0, GPU 1
CPU Socket 1 (16 cores) <-PCIe Gen4-> GPU 2, GPU 3
\end{verbatim}

\begin{itemize}
    \item \textbf{Bandwidth}: PCIe Gen4 x16 provides 31.5 GB/s bidirectional bandwidth per GPU
    \item \textbf{Latency}: Typical PCIe latency of 1-2 microseconds
    \item \textbf{NUMA Effects}: GPUs are attached to specific CPU sockets, requiring careful MPI rank placement to minimize cross-socket traffic
    \item \textbf{GPU-GPU Communication}: NVLink 3.0 provides 600 GB/s bandwidth between GPU pairs (not utilized in our embarrassingly parallel workload)
\end{itemize}

The SETONIX nodes feature AMD's unified Infinity Fabric architecture where each Optimized 3rd Gen EPYC "Trento" CPU (64 cores total) contains 8 chiplets of 8 cores each. Each chiplet is directly connected to one GCD via Infinity Fabric as follows:

\begin{verbatim}
Chiplet 0 (8 cores) <-Infinity Fabric-> GCD 4
Chiplet 1 (8 cores) <-Infinity Fabric-> GCD 5
Chiplet 2 (8 cores) <-Infinity Fabric-> GCD 2
Chiplet 3 (8 cores) <-Infinity Fabric-> GCD 3
Chiplet 4 (8 cores) <-Infinity Fabric-> GCD 6
Chiplet 5 (8 cores) <-Infinity Fabric-> GCD 7
Chiplet 6 (8 cores) <-Infinity Fabric-> GCD 0
Chiplet 7 (8 cores) <-Infinity Fabric-> GCD 1
\end{verbatim}

\begin{itemize}
    \item \textbf{Bandwidth}: Infinity Fabric provides 36-50 GB/s per link (depending on configuration)
    \item \textbf{Latency}: Sub-microsecond latency due to tight integration
    \item \textbf{Coherent Memory}: Unified memory architecture allows coherent access between CPU and GPU memory spaces
    \item \textbf{Natural Affinity}: Each 8-core chiplet has dedicated access to its paired GCD, eliminating NUMA concerns
\end{itemize}

\subsubsection{Performance Implications for Our Workload}

The architectural differences have several implications for the offloaded code segment:

\begin{enumerate}
    \item \textbf{Compute Density}: The AMD MI250X's higher FP64 throughput (47.9 vs 9.7 TFLOPS) directly translates to faster kernel execution for our compute-bound workload, as confirmed by our roofline analysis showing $>80\%$ of peak utilization.
    
    \item \textbf{Memory Bandwidth}: While AMD offers higher memory bandwidth (3276 vs 1555 GB/s), this advantage is less relevant for our compute-bound kernel. However, it does benefit the interpolation table loading phase.
    
    \item \textbf{Host-Device Transfer Overhead}: 
    \begin{itemize}
        \item LEONARDO: PCIe Gen4 limits transfer rates and introduces higher latency
        \item SETONIX: Infinity Fabric's tighter integration reduces transfer overhead, particularly beneficial for smaller, frequent transfers
    \end{itemize}
    
    \item \textbf{Scalability Considerations}:
    \begin{itemize}
        \item LEONARDO: The 4 GPU per node configuration with PCIe can experience bottlenecks when all GPUs simultaneously transfer data
        \item SETONIX: The 8 GCD configuration with dedicated CPU-GPU links provides more consistent performance under full node utilization
    \end{itemize}
    
    \item \textbf{Compilation and Optimization}:
    \begin{itemize}
        \item NVIDIA: Mature compiler toolchain (NVC++ 24.3) with extensive optimization for scientific workloads
        \item AMD: Newer toolchain (amdclang-18) that has rapidly improved but may not yet match NVIDIA's maturity in some optimizations
    \end{itemize}
\end{enumerate}

The combination of superior FP64 performance, better host–device integration, and our code segment compute-bound nature suggests that the AMD platform may exploit its theoretical advantages more effectively. The performance implications of these characteristics are discussed in Section~\ref{Results}. 

Overall, this architectural analysis underscores the importance of portable programming models such as OpenMP, which enable scientific applications to leverage the strengths of diverse GPU architectures without platform-specific optimizations.

\section{{\pinocchio} code}
\label{Pinocchio_GPU_PMT}
{\pinocchio} (\texttt{PIN}pointing \texttt{O}rbit \texttt{C}rossing-\texttt{C}ollapsed \texttt{HI}erarchical \texttt{O}bjects, \citealt{Monaco:2001jg,Monaco:2013qta,Munari:2016aut,Euclid:2025lfa}), is a fast, massively parallel \textbf{C} code, based on LPT and designed to simulate the distribution of DM halos from an initial density field at a given cosmic time.

The {\pinocchio} code operates in three main stages:
\begin{enumerate}
    \item Generation of a linear density contrast field $\delta$ on a uniform grid; this is a Gaussian random process that is connected to the gravitational potential $\phi$ through the Poisson equation: $\nabla^2\phi = \delta$. We refer to this stage as the \texttt{Initialization} module.
    \item Calculation of a ``collapse time'' for each grid point; this is a non-linear function of the eigenvalues of the Hessian of the gravitational potential, $\partial_i\partial_j\phi$. This operation, that is repeated several (10 to 20) times by smoothing the density contrast field over several scales, involves Fast Fourier Transforms (FFTs) to compute the second derivatives of the (smoothed) potential, and a diagonalization of this matrix to obtain the eigenvalues $\lambda_1, \lambda_3, \lambda_3$. These eigenvalues serve as input for a custom solver that estimates the collapse time, described below. The collapse-time calculation is part of a broader module, referred to as \texttt{Fmax}, which also encompasses the previously mentioned FFT operations.
    \item Assembly of collapsed particles into structures, such as halos or filaments, by tracing their hierarchical formation. This phase identifies bound structures and constructs merger trees based on the computed collapse times and a prediction of particle trajectories before collapse. We refer to this stage as the \texttt{Fragmentation} module.
\end{enumerate}

Our GPU porting effort begin from the legacy code publicly available on GitHub at the link \url{https://github.com/pigimonaco/Pinocchio.git}. The legacy version is designed exclusively for multi-core CPU architectures, using a hybrid MPI + OpenMP approach described in~\cite{Monaco:2013qta}. The timing breakdown of this legacy implementation in a full production run is reported in Table~\ref{full_run_pinocchio_timing}. The measurements correspond to the typical configuration described later in Section~\ref{Production_runs}, using 180 compute nodes, each with 16 MPI tasks per node and 2 OpenMP threads per task (32 CPU cores per node), on the NVIDIA platform.
\begin{table}[b!]
\centering
\begin{tabular}{c c c}
\hline
\textbf{Stages} & \textbf{Time [s]} & \textbf{Fraction [\%]} \\
\hline
\texttt{Initialization}  & 103.05  & 3.28 \\
\texttt{Fmax}            & 637.61  & 20.30 \\
\textit{Collapse times (Classic)} & \textit{108.70} & \textit{3.46} \\
\texttt{Fragmentation}    & 2398.90 & 76.38 \\
\hline
Total            & 3140.89 & 100.00 \\
\hline
\end{tabular}
\caption{Timing breakdown of the main stages of a full {\pinocchio} production run. Reported times correspond to wall-clock runtime, and percentages are given relative to the total runtime. 
Collapse times is reported explicitly as part of the \texttt{Fmax} calculation, since this step is the target of our GPU porting.}
\label{full_run_pinocchio_timing}
\end{table}

Despite its modest cost, the collapse time calculation was chosen as the first target for GPU acceleration. Thanks to the modular structure of {\pinocchio}, kernels can be ported incrementally, and the collapse time, computed from the Hessian of the gravitational potential, is the most naturally GPU-portable, being an embarrassingly parallel computation. 

Each mass element (grid cell) is modeled as a homogeneous ellipsoid evolving under its own gravity, with initial conditions given by the Hessian eigenvalues obtained through a standard analytic method. The collapse time of the ellipsoid is then computed with two alternative options. 

In the first option, the evolution of the ellipsoid is approximated using third-order LPT, leading to a non-linear algebraic formulation of the collapse time in terms of the Hessian eigenvalues \citep{Monaco:2001jg}. In the second, the evolution of the ellipsoid is computed without approximations by solving a system of nine Ordinary Differential Equations (ODE) until the density of the ellipsoid diverges \citep[as described in ][]{Song2022}. In the present implementation, this ODE system is integrated on the CPU using a fourth-order Runge–Kutta scheme. The resulting eigenvalues are used to construct interpolation tables, which are subsequently transferred to the GPU and queried by the collapse-time interpolation kernel. 

The latter is mainly used in the case of modified gravity, where the analytic solution based on LPT is not applicable. Since this approach is significantly slower, collapse times are precomputed on a grid of eigenvalues and subsequently interpolated. For computational convenience eigenvalues are sampled in the space defined by the variables $\delta=\Sigma_i\lambda_i$, $x=\lambda_1-\lambda_2$ and $y=\lambda_2-\lambda_3$, with eigenvalues sorted in descending order. 

Since the collapse time function exhibits non-trivial dependence along the $\delta$ direction, requiring non-uniform sampling, the interpolation on this grid is performed using a cubic spline, whereas a simpler (and faster) bi-linear interpolation is performed in the $x$ and $y$ variables.

In both approaches, the code returns the value of the linear growth rate $D(a)$, with $a$ being the scale factor, at the time of collapse $t_{\mathrm{coll}}$. To obtain the corresponding scale factor $a_{\mathrm{coll}}$, the function $D(a)$ must be inverted. To this end, the $D(a)$ function is first computed by solving the ODE governing its evolution, then interpolated with a cubic spline to obtain its inverse $a(D)$. The collapse redshift is finally obtained from the relation $z_{\mathrm{coll}} = 1/a_{\mathrm{coll}} - 1$.
Consequently, spline interpolation is required in both cases.

In the following Sections we will refer to the first approach as the \textit{Classic} collapse time calculation, which only requires the spline interpolation of the inverse collapse time $a(D)$, and to the second as the \textit{Tabulated} collapse time calculation, which additionally involves spline and bilinear interpolation over the precomputed table of eigenvalues. The \textit{Classic} approach corresponds to the configuration adopted in standard production runs, and the timings reported in Table~\ref{full_run_pinocchio_timing} refer to this case.

The GPU porting of the collapse time calculation presented in this work should be viewed in the broader context of an ongoing effort to accelerate the main computational stages of the {\pinocchio} code. In parallel with the GPU porting of the collapse time calculation, substantial effort has been devoted to optimizing the remaining components of the code. The FFT computation has been successfully ported to GPUs using the Highly Efficient FFT for Exascale (heFFTe)~\footnote{\url{https://icl.utk.edu/fft/} library~\citep{tomov2019fftecp}}. At present, the coupling between the FFTs and the subsequent collapse time calculation is limited by host–device communication overhead. Specifically, the quantities produced by the FFTs and required for the collapse time computation (second derivatives of the potential) are currently transferred back to the host and then re-transferred to the device.

Our strategy is to eliminate this overhead by transferring the required data structures to the GPU only once, immediately after the \texttt{Initialization} module, and to subsequently perform both the FFTs and the collapse time calculation entirely on the device. The results will then be transferred back to the host only at the end of the full workflow.

The \texttt{Fragmentation} module, in its current implementation, would not yield significant performance benefits from GPU porting, as it is dominated by conditional branching. Achieving efficient GPU execution would require a substantial re-engineering of the algorithm. For this reason, current efforts are focused on optimizing the CPU implementation. In the longer term, we plan to replace the present \texttt{Fragmentation} module with a machine-learning–based approach trained on N-body simulations. Such an approach is expected to be more amenable to GPU acceleration, owing to its regular computational structure and reduced reliance on conditional branching.

\section{Methodology}
\label{Methodology}
\subsection{Kernel offloading strategy}
\label{Kernel_offloading_Section}

Our effort targets the second step of {\pinocchio}, described in Section~\ref{Pinocchio_GPU_PMT}, specifically the calculation of the collapse times at each grid point (hereafter we refer to it as the target \emph{kernel}, in both its \textit{Classic} and \textit{Tabulated} variants), after the Hessian of the potential has been computed. From an algorithmic perspective, each thread loads the six independent components of the local Hessian and evaluates the collapse time through the interpolation routines described in Section~\ref{Pinocchio_GPU_PMT}, followed by an update of the per-element maxima. The arithmetic workload per grid point is fixed, while memory activity is primarily associated with reading the Hessian components and performing read/modify/write operations on two output arrays. In Roofline terms, this corresponds to a moderate computational intensity, with performance determined by the balance between interpolation arithmetic and memory accesses.

These calculations are inherently independent, making them well-suited for GPU offloading by assigning the computation at each grid point to an individual GPU-thread.

However, the legacy \emph{kernel} logic introduces conditional branches, which can lead to divergent execution paths among GPU threads. It is well established that thread divergence degrades GPU performance by reducing the throughput~\citep{Ryoo_2008, nvidia_cuda_guide_2024}. To mitigate this issue, we eliminate the conditional branches by adopting a masked-based approach (i.e. using  "predicated execution" to avoid branching entirely. The compiler is very good at this, and often a simple if statement can be compiled to use predicated instructions). This ensures that all GPU threads execute the same instructions at the warp/wavefront level, thereby improving computational efficiency. In addition, we reorganized the data layout into a struct-of-arrays (SoA) format, so that consecutive GPU threads access consecutive elements of each array. This improves memory coalescing and further increases throughput~\citep{2025arXiv250609242L}.
\begin{figure}[h!]
    \centering
    \includegraphics[width=0.5\textwidth]{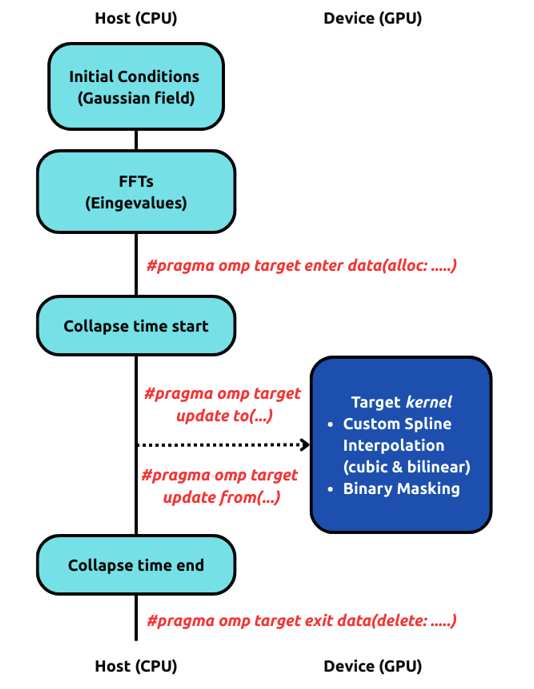}
    \caption{Overview of the GPU offloading strategy used in the {\pinocchio} collapse time kernel. Host-side memory is allocated and selectively transferred to the GPU using OpenMP data mapping directives. The target \emph{kernel}, including the custom GPU-native interpolation routines and binary-masked control flow, runs entirely on the device. Only minimal, necessary data is transferred back to the host, improving performance by reducing memory traffic. 
    }
    \label{Porting_strategy}
\end{figure}

We manage host-device memory transfers explicitly using standard OpenMP data mapping directives as shown in Figure \ref{Porting_strategy}. In particular, we use \texttt{pragma omp target enter data map(alloc:...)} to allocate required quantities on the device at the beginning of the computation, followed by targeted updates using \texttt{pragma omp target update} to transfer only the necessary data between the host and device. This approach avoids unnecessary memory movement and ensures efficient data locality during \emph{kernel} execution. Once the computation is complete, the GPU memory is explicitly released using \texttt{pragma omp target exit data map(delete:...)}.

The computational kernel itself is offloaded using \texttt{pragma omp target teams distribute parallel for} both for initialization of device-resident arrays and for the main per-element loop over grid elements. The \texttt{device(devID)} clause is used to explicitly select the GPU device, and the \texttt{nowait} clause is optionally employed to allow overlap between host and device execution where applicable.

The mapping of iterations to the GPU execution hierarchy (teams and threads) is delegated to the compiler and runtime. We do not explicitly enforce \texttt{num\_teams()} or \texttt{thread\_limit()} constraints. This choice prioritizes portability across GPU architectures, while leaving architecture-specific tuning as potential future work.

At the multi-process level, the mapping of MPI ranks to GPUs is handled by the job scheduler rather than within the application code. On the NVIDIA-based system, MPI ranks are bound to GPUs using the scheduler directive \texttt{--gpu-bind=closest}, which associates each MPI task with the topologically closest GPU and prevents oversubscription. On the AMD system, GPU affinity is managed by the native resource allocation and process placement mechanisms since the equivalent of \texttt{--gpu-bind=closest} is not available yet.

\subsection{Custom interpolation routines implementation}
\label{Custom_interpolation}
A critical part of the \emph{kernel} calculations involves interpolating precomputed quantities required to estimate the collapse time. In the legacy CPU-based implementation, this is handled by GSL library, using cubic spline and bilinear interpolation routines. 

This is necessary because, as discussed Section~\ref{Pinocchio_GPU_PMT}, the linear growth rate $D(a)$ as a function of scale factor $a$ is computed by numerically solving an ODE over a predefined grid. These results are then stored and accessed via cubic spline interpolation, with the spline object constructed once and reused throughout the simulation. 

Additionally, one option for computing collapse times involves tabulating their values on a grid of the eigenvalues $\lambda_i$. The final collapse time is then obtained through interpolation using a mixed scheme: spline interpolation along the dimension with the most complex variation ($\delta$), and bilinear interpolation in the other two dimensions.

For the interpolation of $D(a)$, building the spline on the CPU and transferring it only once (\textit{Classic kernel}) to the GPU is not performance-critical. However, in the second case, where the interpolation table must be updated at every smoothing radius iteration (\textit{Tabulated kernel}), constructing the spline on the host and repeatedly transferring data to the GPU would become a major performance bottleneck. This repeated movement of large data tables would severely impact throughput, making GPU-resident interpolation routines essential for maintaining efficiency.

Since GSL does not support GPU offloading, we developed two custom GPU-native interpolation routines, implemented using OpenMP target constructs. These routines are fully device-resident and designed to avoid host-device transfers during kernel execution. Their numerical behavior closely reproduces the GSL results, as verified in a standalone toy code, while also providing substantial performance improvements. A detailed analysis of accuracy and performance is presented in Section~\ref{Results}.

\subsection{Custom interpolation validation against GSL}
\label{interpolation_validation}
In order to validate the accuracy of the custom GPU-native interpolation routines, we construct a synthetic dataset independent from the main {\pinocchio} workflow.

In particular, we use simple analytical functions with known behavior: a 1D function $y(x) = x^2$ for validating the cubic spline routine, and a 2D function $z(x,y) = x^2 + y^2$ for validating the bilinear interpolation. For the cubic spline test, we use 512 randomly distributed points for building the spline object, consistent with what is typically used within the {\pinocchio} code. For the bilinear case, we construct a 64×64 interpolation table, representative of the structure used in {\pinocchio} runs, where each grid point is associated with a spline of size 128.  The tabulated input points are randomly distributed, reflecting the non-uniform sampling in realistic collapse time tables and highlighting the robustness of our interpolation routines.

In both cases, we generate a set of randomly distributed evaluation points, ranging from $10^2$ to $10^8$, with the upper end corresponding to the number of interpolation calls typically encountered in realistic per-node configurations (e.g., $644^3$ particles).

We then compare the interpolated values produced by the GPU-native routines against those obtained using the original GSL-based CPU implementation. The comparison is quantified by computing the mean residuals across all evaluation points. This procedure allows us to confirm that the numerical differences introduced by the GPU implementation are negligible with respect to the original GSL routines. 

In addition to validating numerical accuracy, we also benchmark the performance of the custom GPU-native interpolation routines against their GSL counterparts as a function of the number of evaluation points. This allows us to estimate the threshold at which GPU acceleration becomes advantageous. 

Validation results and performance tests for the cubic spline and bilinear interpolation routines are presented respectively in Section~\ref{Cubic_spline_validation} and Section~\ref{Bilinear_spline_Validation}. These tests are performed on the NVIDIA platform only, as their purpose is to validate the correctness and assess the relative performance of the GPU-native routines, without requiring cross-platform comparison.

\subsection{Single-node controlled benchmarks}
\label{single_node}
To ensure a meaningful comparison across computing platforms, we perform initial benchmarks using the same number of computational units, hereafter referred to as {\CUs}. Given the heterogeneity of the platforms in terms of the underlying hardware architecture, we define the {\CU} to be $\frac{1}{4}$ of the node, corresponding to eight cores for CPU and one GPU for LEONARDO, and sixteen cores for CPU and two Graphics Compute Dies (GCDs)\footnote{A Graphics Compute Die (GCD) is essentially an independent GPU chip; the MI250X integrates two such dies into a single package.} for SETONIX. This assumption enables a uniform metric for scaling analysis across different configurations. 

The choice of using $\frac{1}{4}$ of the node per {\CU} is motivated by the following considerations:
\begin{itemize}
    \item On LEONARDO, each node hosts four GPUs, and the performance is maximized when each GPU is managed by an MPI process using eight CPU cores (i.e. spawning eight OMP threads);
    \item On SETONIX each GCDs is physically paired with a chiplet of eight CPU cores, supporting a natural 1:8 GCD-to-core mapping.
\end{itemize}
While the current \emph{kernels} implementation allows multiple GPUs to be driven by a single MPI task when needed, this definition of {\CU} provides a practical baseline for fair performance comparisons between CPU and GPU configurations.

All single-node tests on LEONARDO are performed using one computational node, with up to 4 MPI processes (corresponding to 32 cores for CPU runs and 4 GPUs for accelerator runs). All single-node tests on SETONIX are performed using one computational node, with up to 8 MPI processes (corresponding to 64 cores for CPU runs and 8 GCDs (4 GPUs) for accelerator runs). This setup ensures that performance measurements are not affected by inter-node communication.

Single-node benchmarks are performed for both the \textit{Classic} and \textit{Tabulated} collapse time \emph{kernels}. In all cases, the \emph{kernel} time-to-solution is defined as the wall-clock time required for the actual computation. For GPU runs, this includes both the device execution time and the overhead from host-device data transfers, providing a realistic assessment of end-to-end performance.

For a multi-node weak and strong scaling analysis analogous to the single-node benchmarks presented here, we refer the reader to the companion work presented in~\cite{LACOPO2026101060}, where consistent speedup trends, similar to those discussed later in this work, are reported.

\subsection{Production run setup}
\label{Production_runs}
While the single-node benchmarks focus on a controlled environment for platform comparison, production runs adopt a different configuration optimized for the overall application performance. Specifically, we increase the number of MPI tasks per node and reduce the number of OpenMP threads per task. This configuration better matches the scaling characteristics of other parts of {\pinocchio}, such as the FFT computations and the \texttt{Fragmentation}, which benefit from a finer-grained domain decomposition. 

In these tests, we run with 16 MPI tasks per (4 per GPU, illustrating the flexibility of the MPI–GPU mapping while preserving correctness) node, each using 2 OpenMP threads, for a total of 2880 MPI tasks across 180 compute nodes (using 32 cores per node). 

These production-scale benchmarks are performed exclusively on the NVIDIA platform, where such a large allocation of resources is available, and are carried out with the \textit{Classic kernel}, which corresponds to the standard configuration in production runs not involving modified gravity (which lies beyond the scope of this work).

To ensure that the GPU offloading does not impact the scientific correctness of the results, we compare the distribution of collapsed particle, as a function of redshift z, produced by the GPU and legacy CPU implementations. This quantity is a direct output of the code and encapsulates the physics modeled by the \emph{kernel}.

\section{Results}
\label{Results}
In this section, we present the results of the GPU porting of the {\pinocchio} \emph{kernels}, focusing on both numerical accuracy and performance. We begin by validating the custom GPU-native interpolation routines through a comparison with the original GSL-based implementation, using a standalone toy code developed out of {\pinocchio}. We then present the single-node performance analysis, comparing the GPU \emph{kernels} and CPU legacy implementations presented in Section~\ref{Pinocchio_GPU_PMT}, across the two architectures described in Section~\ref{Computing_platforms}, while varying the number of {\CUs} as explained in Section~\ref{single_node}. This is followed by a roofline analysis, performed for the \textit{Classic kernel} only since it represents the production-relevant case, which provides insight into the computational efficiency relative to the architectural performance ceilings. Finally, we analyze the impact of the GPU \emph{Classic kernel} in a full production scenario, highlighting both the performance improvements and the scientific consistency of the results, following the resources setup discussed in Section~\ref{Production_runs}. 

For all benchmarks, performance results are based on a single run. However, the repeated evaluation of collapse times across 10–20 smoothing radii, as explained in Section~\ref{Pinocchio_GPU_PMT}, effectively samples variability, yielding consistent timings. 

All NVIDIA benchmarks were compiled using the \texttt{NVC/NVC++} compilers (v24.3) with the following optimization and offloading flags:
\texttt{-O3 -fast -mp=multicore,gpu -target=gpu -gpu=ccnative}.
For the AMD platform, compilation was performed using the \texttt{amdclang-18} compiler with the following flags:
\texttt{-O3} \texttt{-fopenmp} \texttt{-offload-arch=gfx90a} \texttt{-mtune=native}.

\subsection{Cubic spline validation}
\label{Cubic_spline_validation}
\begin{figure*}[t]
    \centering
    \includegraphics[width=0.75\textwidth]{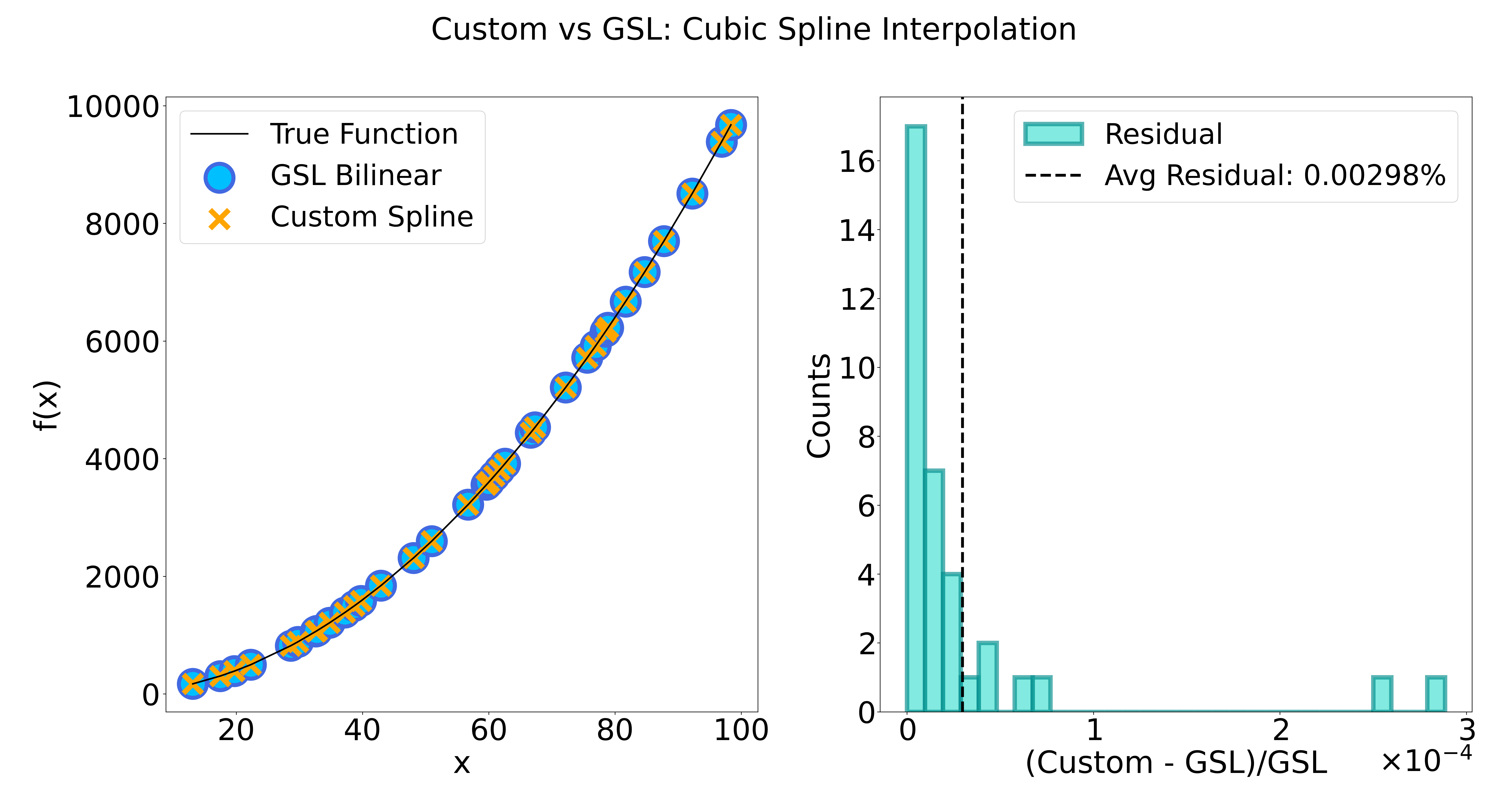}
    \caption{Left panel displays the comparison between the GPU-custom cubic spline interpolation routine, the GSL implementation, and the analytical 1D reference function described in Section~\ref{interpolation_validation}. A total of only 35 evaluation points are shown for clarity. Right panel shows the histogram of the residuals between the GPU-custom and GSL interpolations, aggregated over the 35 evaluation points. Residuals are shown in absolute units, while the dashed line indicates the average residual expressed as a percentage.
    }
    \label{Cubic_spline_vs_GSL}
\end{figure*}
\begin{figure*}[h!]
    \centering
    \includegraphics[width=0.75\textwidth]{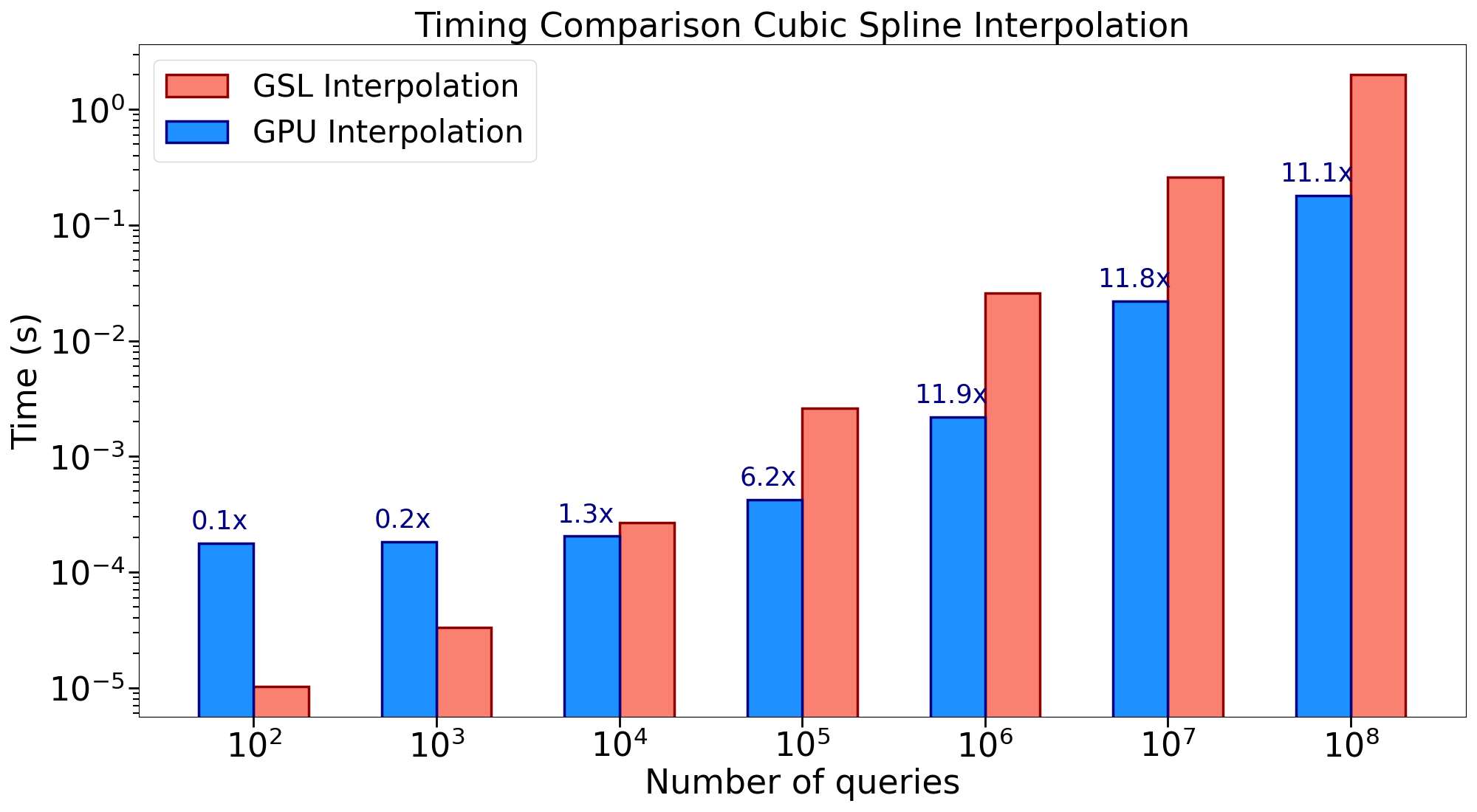}
    \caption{Comparison of interpolation wall-time between the GPU-native cubic spline routine and the GSL implementation, as a function of the number of evaluation points. GPU timings include memory transfers between the host and device. Speedup values relative to the GSL implementation are reported above each GPU bar. CPU results correspond to one MPI task with 32 OpenMP threads (full node), while GPU results use one MPI task driving 4 GPUs.}
    \label{Cubic_spline_vs_GSL_timing}
\end{figure*}

To evaluate the accuracy and performance of the cubic spline interpolation, we test the GPU-native routine against the GSL implementation using the synthetic 1D function described in Section~\ref{interpolation_validation}.

Figure~\ref{Cubic_spline_vs_GSL} summarizes the accuracy results. The left panel displays the interpolated values obtained from both the GPU-native and GSL-based routines, alongside with the analytical reference function. For visual clarity, only 35 randomly distributed evaluation points are displayed. As shown, the GPU custom implementation closely follows both the analytical reference and the well-established GSL results, confirming the numerical accuracy of our custom routine. 

The right panel shows the histogram of the residuals between the GPU and GSL interpolations. The dashed line marks the average residual, which remains at the level of sub-percent level ($\sim0.003\%$), indicating excellent agreement between the two methods. A slight systematic overestimation is observed, with residuals consistently positive across the evaluation points. This mild systematic, likely introduced by the numerical treatment of second derivatives in the custom GPU implementation, remains well below the percent level. As will be shown in Section~\ref{Production_run_results}, the impact of these deviations on the final scientific output is negligible.
\begin{figure*}[t]
    \centering
    \includegraphics[width=0.75\textwidth]{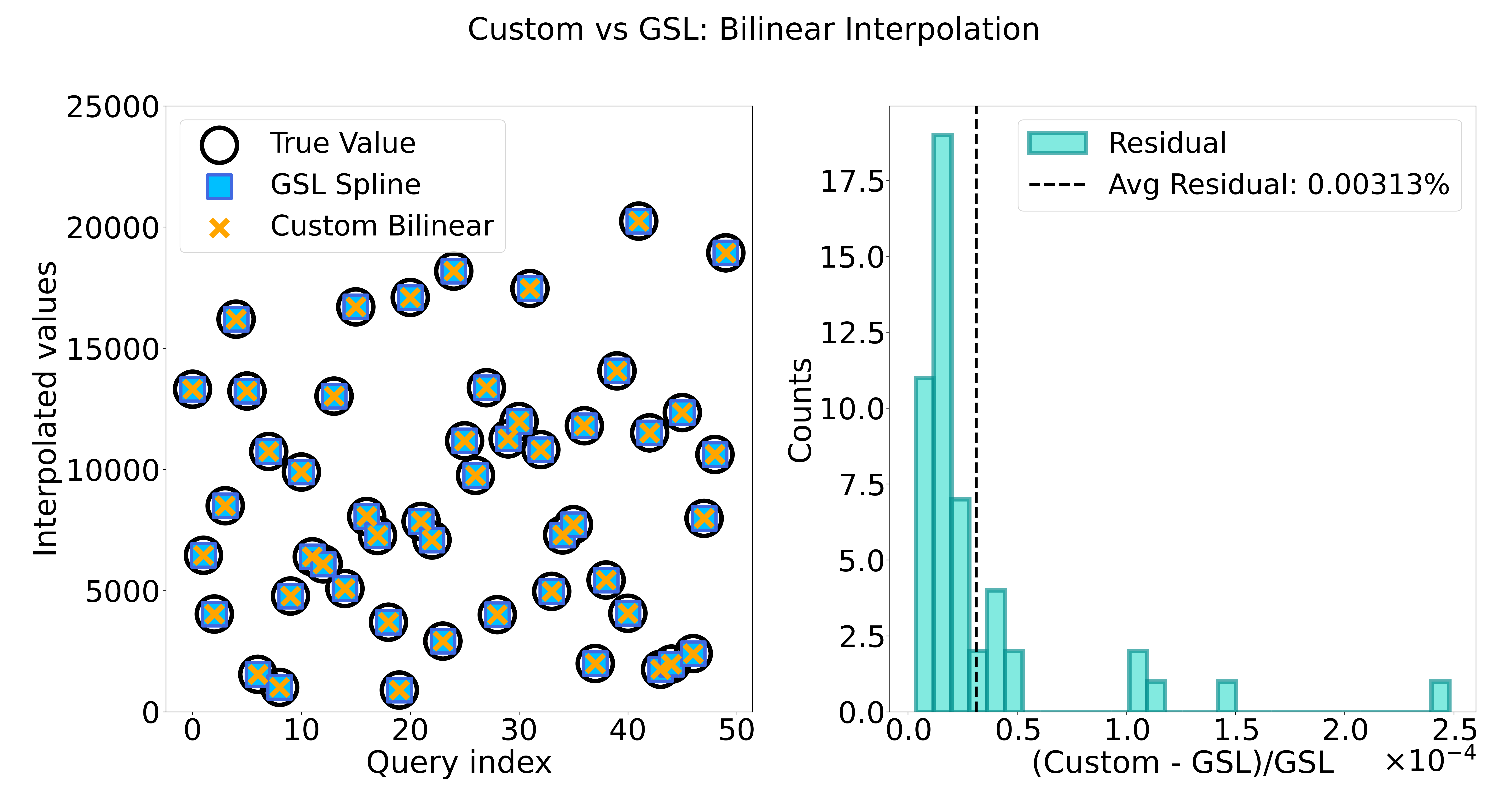}
    \caption{Same as Figure~\ref{Cubic_spline_vs_GSL} but for the bilinear interpolation routines. In this case 50 evaluation points are shown in the left panel.
    }
    \label{Bilinear_interpolation_vs_GSL}
\end{figure*}

In addition of validating numerical accuracy, we evaluate the performance of the custom GPU-native cubic spline interpolation routine against the GSL implementation in Figure~\ref{Cubic_spline_vs_GSL_timing}. For this standalone toy-code benchmark, the full node was used. The CPU run employed one MPI task with 32 OpenMP threads (32 CPU cores), while the GPU run used one MPI task coupled with the 4 GPUs available on the node.

The GPU-custom interpolation is shown in blue, while the GSL-based implementation is shown in red. At small numbers of interpolation queries, GPU performance is reduced primarily due to insufficient parallelism and low occupancy, which prevent the device from fully utilizing its compute resources. 
\begin{table}[h!]
\centering
\resizebox{\linewidth}{!}{%
\begin{tabular}{lccccc}
\hline
Method & Mean [s] & Std [s] & Median [s] & Min--Max [s] & RSD [\%] \\
\hline
Custom GPU & 0.181 & 0.087 & 0.161 & 0.161--0.436 & 45.98 \\
GSL CPU    & 2.003 & 0.070 & 1.980 & 1.980--2.202 & 3.50 \\
\hline
\end{tabular}
}
\caption{Run-to-run timing statistics over 10 repetitions for the largest workload ($10^8$ interpolation queries) for the cubic spline interpolation.}
\label{tab:timing_variability}
\end{table}
In this regime, \textit{kernel} launch overheads and latency cannot be effectively amortized. However, as the number of evaluation points increases, the GPU routine scales more efficiently, becoming faster beyond approximately $10^4$ evaluations and reaching a speed up factor of $\sim 11- 12\times$ relative to the GSL.

The GPU timings include memory transfers between the host and device, providing a realistic estimate of performance in practical settings. To characterize performance robustness, each interpolation benchmark was repeated 10 times on the same dataset. In Table~\ref{tab:timing_variability} we provide a detailed timing statistics for the largest tested workload ($10^8$ interpolation queries). In particular, we report the mean, the standard deviation, and the median together with the min-max range, and the relative standard deviation (RSD), defined as the standard deviation normalized by the mean execution time and expressed as a percentage. The CPU implementation exhibits low run-to-run variability, while the GPU measurements show larger dispersion due to runtime and data-transfer effects. The median remains close to the minimum execution time, indicating that occasional slower runs do not alter the observed performance trend.

Such performance gains become particularly relevant in \pinocchio\ runs, where collapse times must be evaluated for billions of particles across multiple smoothing radii. Despite the observed run-to-run variability in the GPU timings, the overall speedup with respect to the CPU implementation is preserved, indicating that the performance improvement is robust for production workloads.

\subsection{Bilinear validation}
\label{Bilinear_spline_Validation}
\begin{table}[h!]
\centering
\resizebox{\linewidth}{!}{%
\begin{tabular}{lccccc}
\hline
Method & Mean [s] & Std [s] & Median [s] & Min--Max [s] & RSD [\%] \\
\hline
Custom GPU & 0.499 & 0.099 & 0.467 & 0.467--0.781 & 19.87 \\
GSL CPU    & 18.058 & 0.114 & 18.022 & 18.021--18.383 & 0.63 \\
\hline
\end{tabular}
}
\caption{Same as Table~\ref{tab:timing_variability} but for the bilinear interpolation routine.}
\label{tab:timing_variability_alt}
\end{table}
Following the same approach adopted for the cubic spline interpolation described in Section~\ref{Cubic_spline_validation}, we assess the numerical accuracy and performance of the GPU-native bilinear interpolation by comparing its output with the GSL implementation across a set of randomly distributed evaluation points.

The left panel of Figure~\ref{Bilinear_interpolation_vs_GSL} shows the interpolated values from both methods, compared against the ground truth values derived from the analytical function, demonstrating visually consistent results. The right panel presents the histogram of the residuals between the two methods. The numerical accuracy ($\sim0.003\%$) and the mild overestimation trend is consistent with what is observed for the cubic spline interpolation.

\begin{figure*}[h!]
    \centering
    \includegraphics[width=0.75\textwidth]{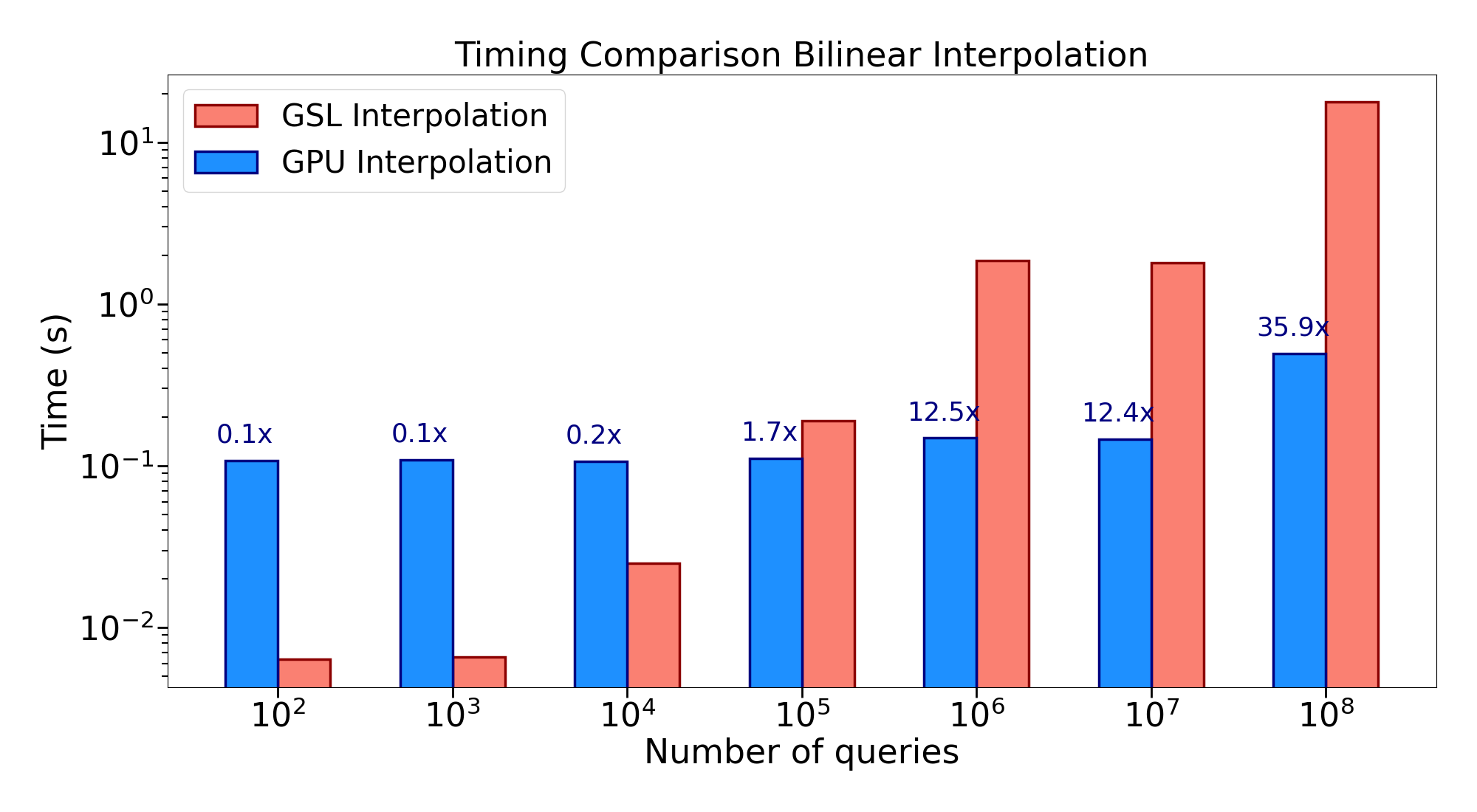}
    \caption{Same as Figure~\ref{Cubic_spline_vs_GSL_timing} but for the bilinear interpolation routines.
    }
    \label{Bilinear_vs_GSL_timing}
\end{figure*}

Figure~\ref{Bilinear_vs_GSL_timing} shows the interpolation performance comparison, following the same setup, evaluation strategy, and  resource configuration described for the cubic spline in Figure~\ref{Cubic_spline_vs_GSL_timing}.
Interestingly, in contrast to the cubic spline case, the GPU-custom bilinear interpolation routine exhibits nearly constant execution time across varying evaluation size. This behavior suggests efficient parallel utilization and early kernel saturation, where the GPU is able to fully occupy its compute units even at modest workloads. However, due to the higher per-query complexity of the bilinear interpolation, involving four spline evaluations per call and more scattered memory access patterns, the speedup relative to the GSL implementation becomes evident only at larger evaluation point ($\geq 10^6$). This scaling behavior illustrates the efficiency of the GPU implementation in scenarios involving heavy interpolation workloads. 

For completeness, in Table~\ref{tab:timing_variability_alt} we report the same timing analysis as in Table~\ref{tab:timing_variability}, but for the custom bilinear interpolation routine. 
The smaller GPU dispersion observed in this case reflects the different balance between runtime overhead and arithmetic workload.  Kernels with longer execution times exhibit reduced relative variability because fixed runtime effects (kernel launch latency, synchronization, and data-transfer overheads) represent a smaller fraction of the total runtime.

\subsection{Single-node speedup comparison}
\label{single_node_test}
\begin{figure*}[t]
    \centering
    \includegraphics[width=0.65\textwidth]{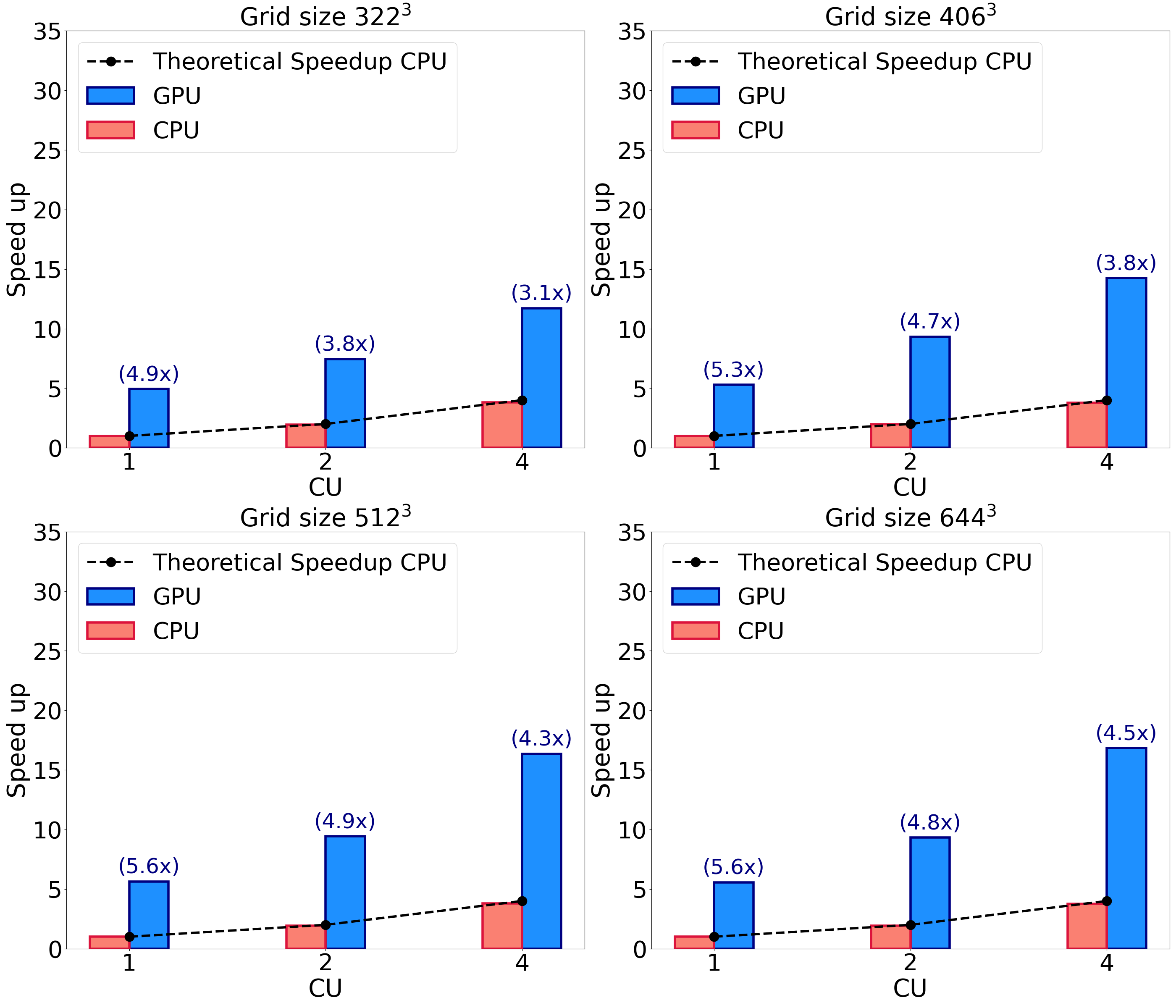}
    \caption{Speedup in strong-scaling tests for NVIDIA platform for different grid sizes, using the \textit{Classic kernel}. The plots display performance as grid resolution increases from the upper-left to the lower-right. Bar heights represent the normalized speedup relative to the 1 CPU {\CU} baseline. Labels above GPU bars indicate the speedup of the GPU implementation relative to the CPU implementation at the same number of {\CU}.
    }
    \label{NVIDIA_perfomance_node}
\end{figure*}
\begin{figure*}[h!]
    \centering
    \includegraphics[width=0.65\textwidth]{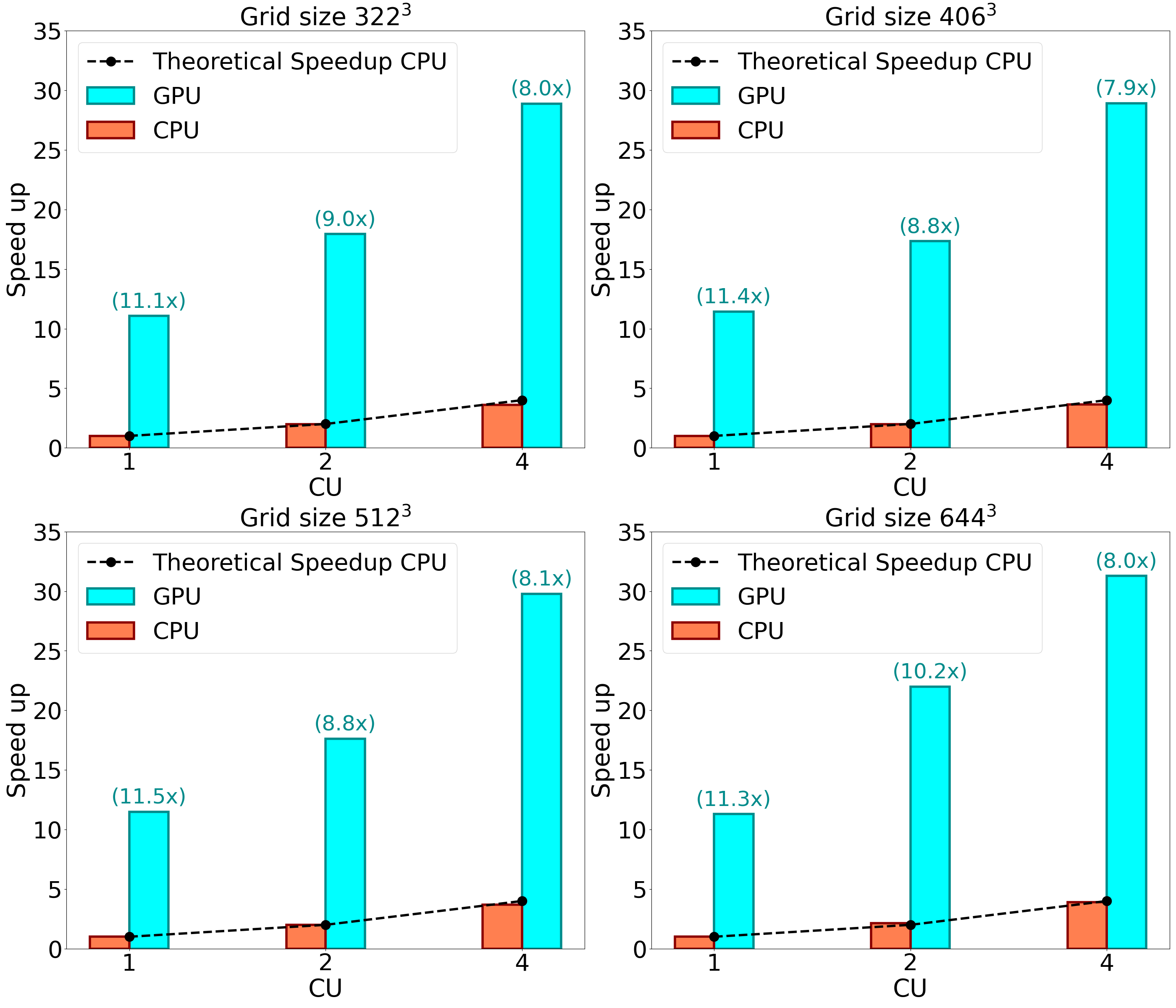}
    \caption{Same as Figure~\ref{NVIDIA_perfomance_node} but for the AMD platform.}
    \label{AMD_perfomance_node}
\end{figure*}
We begin our performance analysis of the {\pinocchio} \emph{kernels} with strong-scaling benchmarks on a single node, comparing GPU and CPU implementations across different platforms and grid sizes. This analysis is performed for both collapse time computation methods described in Section~\ref{Pinocchio_GPU_PMT}, namely the \textit{Classic} and \textit{Tabulated kernels}. 

Unless otherwise stated, all CPU results reported in this section use the following configurations. On the NVIDIA platform, each CPU \CU\ corresponds to one MPI rank coupled with 8 OpenMP threads, for a total of 8 physical CPU cores per \CU. The three configurations therefore use 1, 2, and 4 MPI ranks with 8 OpenMP threads per rank, corresponding to 8, 16, and 32 CPU cores, respectively. On the AMD platform, each CPU \CU\ corresponds to one MPI rank coupled with 16 OpenMP threads, for a total of 16 physical CPU cores per \CU. The three configurations therefore use 16, 32, and 64 CPU cores, respectively. A detailed discussion of the resource allocation strategy is provided in Section~\ref{single_node}.

CPU speedup is defined as the ratio between the time-to-solution using 1, 2, and 4 CPU {\CU}s and that using a single CPU \CU. GPU results are similarly normalized to the single CPU \CU\ way for visual comparison, but labels above GPU bars indicate the speedup relative to the CPU implementation using the same number of {\CU}s. This enables a direct, per-{\CU} comparison of GPU versus CPU performance at each scale.

Figure~\ref{NVIDIA_perfomance_node} and~\ref{AMD_perfomance_node} illustrates the results for NVIDIA and AMD platforms respectively, with increasing grid resolution moving from the upper-left to the lower-right for the \textit{Classic kernel}. For completeness, the CPU performance on both platforms is included, alongside with the expected theoretical speedup for CPU calculations. The results validate the parallel efficiency of the \textit{Classic} kernel computations. CPU speedup closely follows the theoretical expectation across all grid sizes, confirming that the workload scales efficiently and evenly across the available {\CU}. Additionally, the GPU implementation consistently achieves a speedup at least 4$\times$ higher than the CPU across all {\CU}s. This consistent gain confirms the suitability of GPUs in handling massively parallel workloads such as the collapse time computation. Notably, the advantage of GPU offloading becomes more pronounced at higher grid resolutions ($512^3$ and $644^3$), where the increased computational load leads to better GPU utilization. At smaller grid sizes, the workload may be insufficient to fully saturate the GPU, which limits the achievable speedup. 
\begin{figure*}[t]
    \centering
    \includegraphics[width=0.65\textwidth]{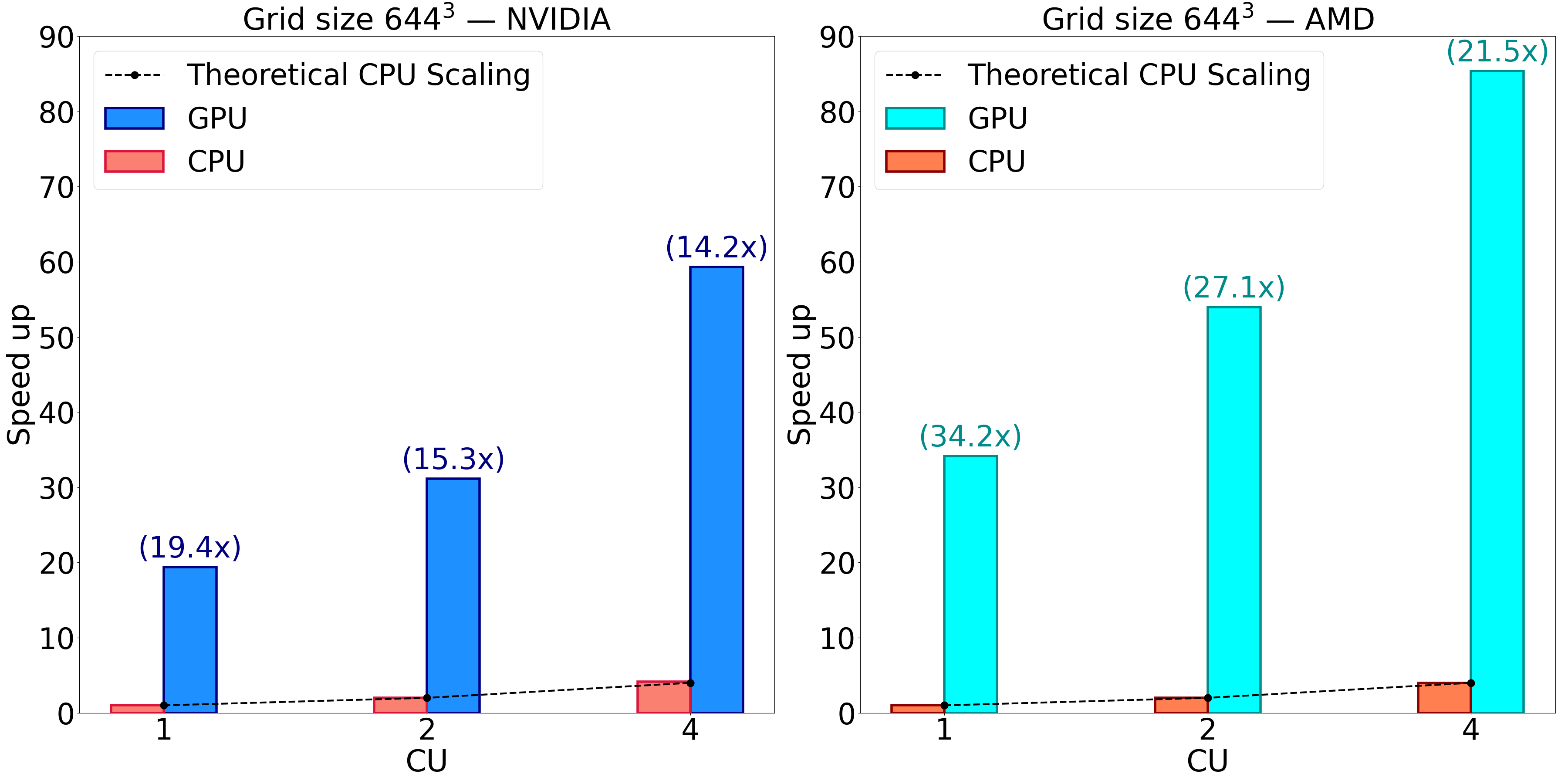}
    \caption{Same as Figure~\ref{NVIDIA_perfomance_node}, but using the \textit{Tabulated kernel}. Results are shown for the NVIDIA platform (left) and AMD platform (right), and refer exclusively to the $644^3$ grid size.}
    \label{Bilinear_performance}
\end{figure*}
\begin{figure*}[h!]
    \centering
    \includegraphics[width=0.65\textwidth]{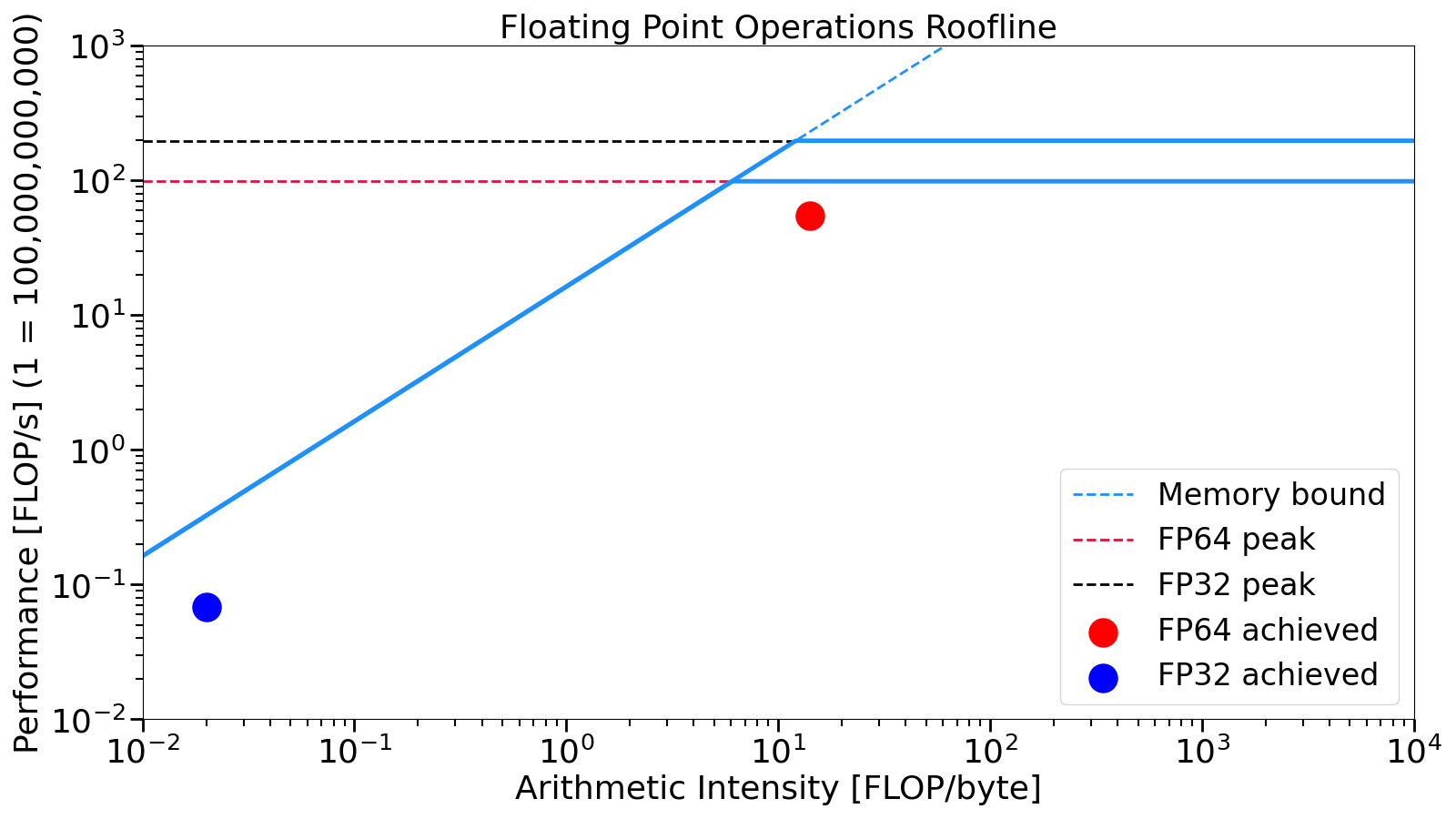}
    \caption{Roofline analysis of the GPU-accelerated \emph{Classic kernel} on the NVIDIA platform. The \textit{kernel} achieves over 80\% of the theoretical FP64 peak performance, operating in the compute-bound regime near the performance plateau (red dot).
    }
    \label{Roofline_analysis}
\end{figure*}

For this reason, in the case of the \textit{Tabulated kernel}, we report performance results only for the $644^3$ grid, where the computational intensity is sufficient to highlight the scaling behavior and hardware differences. The results are shown in Figure~\ref{Bilinear_performance}, where the left panel presents the NVIDIA case, and the right panel corresponds to the AMD platform.
The larger GPU speedup observed for the \textit{Tabulated kernel} (up to 4$\times$ higher than in the \textit{Classic kernel}) is primarily due to the absence of OpenMP parallelization in the corresponding CPU implementation. While both tests use the same number of CPU cores per CU, the \textit{Classic kernel} benefits from hybrid MPI + OpenMP parallelism, whereas the \textit{Tabulated kernel} relies solely on MPI. As a result, the CPU baseline is less optimized in the latter, amplifying the apparent GPU performance gain. It is also worth noting that the \textit{Tabulated} approach is mainly used in runs involving modified gravity, as explained in Section~\ref{Pinocchio_GPU_PMT}, which are beyond the scope of this work. For this reason, our large-scale production tests in Section~\ref{Production_run_results} focus on the \textit{Classic} variant only.

Notably in both approaches cases, we also observe that the speedup achieved on the AMD platform is roughly twice that of the NVIDIA platform. As discussed in Section~\ref{GPU_architectures}, the combination of approximately 5$\times$ difference in FP64 peak performance (47.9 vs 9.7 TFLOPS) between AMD GCDs and NVIDIA A100s, better host-device integration, and our kernel's compute-bound nature allows the AMD platform to more effectively utilize its theoretical advantages.

These findings confirm that our GPU implementations deliver both substantial performance improvements and excellent portability. The consistent scaling behavior across NVIDIA and AMD platforms highlights the robustness of our OpenMP offloading approach for heterogeneous computing environments.

\subsection{Roofline analysis}
\label{Roofline_test}

To quantitatively assess the computational efficiency of our GPU porting strategy, we perform a roofline analysis on the NVIDIA platform using NVIDIA Nsight Compute tool\footnote{\url{https://developer.nvidia.com/nsight-compute}}. This allows us to evaluate how close the GPU \emph{Classic kernel} operates to the theoretical hardware performance limits and to confirm that thread divergence has been effectively mitigated by the masked-based implementation described in Section~\ref{Methodology}.

This analysis was performed on the NVIDIA platform only, and for the \textit{Classic kernel}, which reflects the setup adopted in our production simulation runs. The \textit{Tabulated kernel} is also GPU-accelerated and structurally similar to the \textit{Classic} one; however, it is currently used mainly in exploratory configurations and is not part of the standard \pinocchio\ production workflow. Since the purpose of this Section is to characterize performance in representative production conditions, we restrict the detailed profiling analysis to the \textit{Classic} \textit{kernel}, which constitutes the default collapse-time computation path. Support for the \textit{Tabulated kernel} is provided to enable extended use cases, and a dedicated assessment of its performance characteristics is left to future work. Nevertheless, similar behavior is expected, as the \textit{Tabulated kernel} builds upon the \textit{Classic} one with the addition of a bilinear interpolation routine, which is lightweight and free of conditional branching.
\begin{table*}[t]
\centering
\begin{tabularx}{0.6\textwidth}{c XX XX}
\hline
& \multicolumn{2}{c}{\textbf{CPU}} & \multicolumn{2}{c}{\textbf{GPU}} \\
\textbf{Stages} & \textbf{Time [s]} & \textbf{[\%]} & \textbf{Time [s]} & \textbf{[\%]} \\
\hline
\texttt{Initialization}  & 103.05  & 3.28  & 86.66  & 2.88 \\
\texttt{Fmax}            & 637.61  & 20.30 & 550.53 & 18.29 \\
\textit{Collapse times (Classic)} & \textit{108.70} & \textit{3.46} & \textit{18.30} & \textit{0.61} \\
\texttt{Fragmentation}   & 2398.90 & 76.38 & 2371.21 & 78.78 \\
\hline
\textbf{Total}           & 3140.89 & 100.00 & 3009.84 & 100.00 \\
\hline
\end{tabularx}
\caption{Timing breakdown of the main stages of a full {\pinocchio} production run for CPU and GPU executions. 
Reported times correspond to wall-clock runtime, and percentages are relative to the total runtime. 
Collapse times are shown explicitly as part of the \texttt{Fmax} calculation, since this step is the target of the GPU porting.}
\label{full_run_pinocchio_timing_gpu}
\end{table*}

Figure~\ref{Roofline_analysis} illustrates the results. The \textit{Classic kernel} achieves over 80\% of the theoretical peak performance in FP64, indicating efficient use of available GPU resources. The measured arithmetic intensity places the \textit{kernel} well within the compute-bound regime (red dot in Figure~\ref{Roofline_analysis}), close to the performance plateau, confirming that the GPU is effectively saturated and that the implementation makes good use of available compute throughput. Arithmetic intensity is measured using NVIDIA Nsight Compute hardware performance counters collected during \textit{kernel} execution. The profiler reports the number of executed floating-point operations and the volume of data transferred from global memory; the arithmetic intensity is computed as their ratio, following the standard roofline definition. No manual estimation of FLOPs or memory traffic is performed.

We stress that this compute-bound characterization applies strictly to on-device \textit{kernel} execution; the dominance of host–device data transfers discussed in Section~\ref{data_movement} concerns the overall workflow and does not contradict the roofline classification of the \textit{kernel} itself. This high efficiency aligns with the speedup trends observed in our timing benchmarks and further validates the quality of the GPU porting.

\subsection{Production run}
\label{Production_run_results}

While benchmarks presented in Section~\ref{single_node_test} focused on controlled single-node conditions to ensure platform comparability, production-scale runs involve a configuration tailored to the overall performance of the entire~{\pinocchio} application. In these cases, domain decomposition, FFT scalability and the halo construction workflow play a critical role, motivating a shift toward a more MPI-intensive configuration. To reflect realistic usage scenarios, we adopt the hybrid parallel configuration described in Section~\ref{Production_runs}. These production-scale tests were performed on the NVIDIA platform, which currently provides access to a sufficient number of compute nodes required for large-scale simulations.

We evaluate the GPU-accelerated version of the \emph{Classic kernel} within the full production workflow, measuring the time-to-solution, as detailed in Table~\ref{full_run_pinocchio_timing_gpu}. The collapse-time stage is accelerated by approximately a factor of
$\sim6\times$ (108 s on CPU versus $\sim18$ s on GPU), consistent with the behavior observed in the controlled single-node benchmark(Section~\ref{single_node_test}). The CPU baseline corresponds to the production configuration described in Section~\ref{Production_runs}, using 16 MPI tasks per node with 2 OpenMP threads per task, corresponding to 32 CPU cores per node across 180 compute nodes. The GPU configuration uses the same number of nodes, with 4 GPUs per node and 16 MPI tasks per node, for a total of 720 GPUs.
\begin{figure*}[t]
    \centering
    \includegraphics[width=0.65\textwidth]{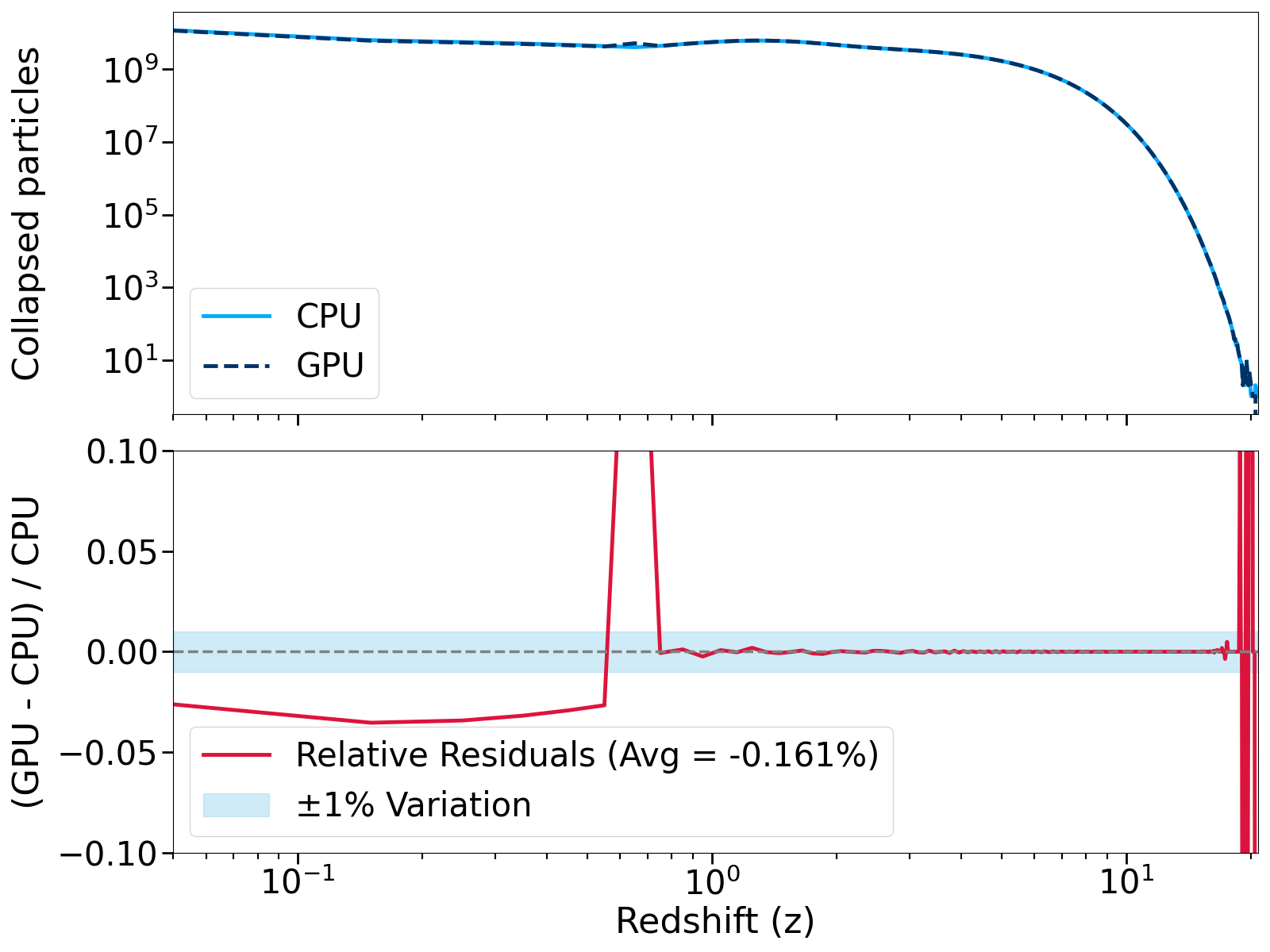}
    \caption{Comparison of the number of collapsed particles per redshift bin  obtained using the CPU and GPU implementations of the \textit{Classic kernel}. The upper panel shows the total number of collapsed particles as a function of redshift z, while the lower panel presents the relative residuals between the GPU and CPU results. The shaded region highlights the $\pm1\%$ deviation band and the dashed lined indicates the level of perfect agreement.
    }
    \label{FMAX_cubic_spline}
\end{figure*}

This confirms that the benefits of GPU offloading persist under realistic, large-scale workloads, validating the scalability of the implementation. In practical terms, this translates into a net saving of approximately 90-100 seconds per simulation,  compatible with the run-to-run variability. Given that typical cosmological campaigns involve thousands of {\pinocchio} runs, this results in a cumulative saving of over 160000 Standard-h, a significant reduction in computational cost that enables more efficient use of HPC resources. Such a saving also implies a non-negligible reduction in energy consumption; however, a quantitative assessment of this aspect requires dedicated analysis, which is shown in~\cite{LACOPO2026101060} companion work.

In addition to performance, we validate the scientific consistency of the GPU implementation by comparing the distribution of collapsed particles, as a function of redshift z, obtained in the GPU and CPU runs. 

As shown in Figure~\ref{FMAX_cubic_spline}, the two distributions are in excellent agreement, confirming that the ported \emph{Classic kernel} accurately reproduces the expected collapse statistics. The relative residuals remain within $\pm1\%$ across most of the redshift range, demonstrating the robustness of the GPU interpolation scheme. For $z \gtrsim 0.6$,  the agreement is consistently better than $1\%$. At $z \lesssim 0.5$,  the GPU version exhibits a small systematic underestimation in the number of collapsed particles, reaching up to $\sim 2.5-3 \%$. Interestingly, this trend is interrupted by a localized overestimation around $z \sim 0.5-0.6$, where the GPU result overshoots the CPU reference. This anomaly may be attributed to a subtle interpolation artifact in the inverse growing mode near the transition between rapid and slow growth, where the cubic spline response becomes more sensitive to local curvature. At high redshifts ($z \gtrsim 10$), discrepancies appear large in relative terms but are statistically irrelevant, as they involve an extremely small fraction of collapsing particles at this epoch. Overall, the results confirm the scientific fidelity of the GPU implementation, with only minimal deviations from the CPU reference. 

As will be shown in Appendix~\ref{HMF_comparison}, where we compare the resulting numerical halo mass functions (HMFs), this overall mild underestimation of number of collapse particle ($\sim 0.16\%$) is effectively absorbed by the \texttt{Fragmentation} algorithm. The resulting HMFs from the CPU and GPU versions are in excellent agreement across all mass range.  Furthermore, as shown in~\cite{Munari:2016aut}, the intrinsic deviation of \pinocchio\ relative to full N-body simulations is typically in the range of $5-10\%$ for the HMF. In this context,  the overall underestimation of collapse times at the level of $\sim0.16\%$, as illustrated in Figure~\ref{FMAX_cubic_spline} is well within acceptable tolerances and has no significant impact on the physical predictions.

\subsection{Data movement costs}
\label{data_movement}
In Sections~\ref{single_node_test} and~\ref{Production_run_results} we focused primarily on runtime speedup. Here, we provide a more detailed breakdown of the collapse-time stage for both the \textit{Classic} and \textit{Tabulated} \textit{kernels}, clarifying the relative contributions of computation and data movement in the GPU implementation relative to the CPU baseline. Unless otherwise stated, the breakdown reported below refers to the NVIDIA platform.

In the single-node configuration, with 4 \CUs, and for the largest grid resolution (grid-size $644^3$), the collapse-time computation accounts for $\sim2.5\%$ for the \textit{Classic kernel} and $\sim4.5\%$ for the \textit{Tabulated kernel} of total runtime in the CPU implementation. After GPU offloading, this contribution decreases to approximately $\sim0.3\%$ in both cases.

A similar trend is observed in the production-run configuration, where the collapse-time stage decreases from $\sim3.5\%$ of the total runtime on CPU, to $\sim0.6\%$ on GPU (see Table~\ref{full_run_pinocchio_timing_gpu}). These results demonstrate that GPU acceleration effectively reduces the relative weight of the collapse-time stage within the overall workflow. Once ported to the GPU, however, the residual cost of this stage is dominated by host–device data transfers rather than on-device computation.

In the single-node configuration, GPU computation accounts for only $\sim0.01$–$0.02\%$ of total runtime, while host–device transfers account for $\sim0.32$–$0.33\%$. An analogous behavior is observed in the production run, where computation contributes $\sim0.02\%$ and data transfer $\sim0.6\%$. This indicates that further performance gains will primarily depend on reducing data movement and improving data residency on the device.

A similar behavior is observed on the AMD platform in the single-node benchmark, with 4 \CUs\ for both \textit{kernels}  and the same grid-size of $644^3$: after GPU porting the collapse-time stage accounts for $\sim0.27\%$ of the runtime, of which only $\sim0.03\%$ corresponds to GPU computation and $\sim0.24\%$ to memory transfers. This confirms that the performance limitation is not architecture-specific but primarily due to host–device data movement. As anticipated in Section~\ref{GPU_architectures}, host-device communication is slightly faster due to tighter host-device integration. We stress that the production run has not been tested on the AMD platform because we did not have access to a sufficient number of compute nodes required for large-scale simulations.

\section{Conclusion}
\label{Conclusion}
In this work, we presented the porting and optimization of a  specific segment  within the \pinocchio\ cosmological simulation code to GPUs, using OpenMP target directives as a portable, directive-based programming model. Our effort focused specifically on the collapse time calculation, an embarrassingly parallel segment of the simulation pipeline ideally suited for GPU acceleration, and which in {\pinocchio} can be computed using either the \textit{Classic} or the \textit{Tabulated kernels}.

To ensure full GPU offloading and architectural portability, we developed custom GPU-native implementations of both cubic spline and bilinear interpolation routines, which constitute the computational core of the aforementioned \textit{kernels}. These implementations successfully eliminate reliance on the GNU Scientific Library (GSL), which lacks GPU support, and were written entirely with OpenMP pragma-based directives to ensure compatibility across both NVIDIA and AMD GPU architectures. Validation tests, using a standalone toy code, confirm that the numerical results from the GPU-native interpolation routines maintain high fidelity, with residuals typically at the sub-percent level ($\sim0.003\%$) compared to the original GSL-based CPU versions (see Section~\ref{Cubic_spline_validation} and Section~\ref{Bilinear_spline_Validation}). Performance benchmarks further demonstrate that our GPU implementations outperform their CPU counterparts once a sufficient number of evaluations points is reached, achieving up to 12$\times$ speedup for both interpolation routines and showing significant scalability benefits in production-scale use cases.

Strong scaling tests, of the offloaded \emph{kernels} within the \pinocchio\ code, across both NVIDIA (LEONARDO) and AMD (SETONIX) supercomputing platforms consistently show GPU speedups exceeding 4$\times$ relative to CPU-only implementations, with even larger gains observed for the \textit{Tabulated kernel} due to the lack of OpenMP parallelization in the corresponding CPU baseline (see Section~\ref{single_node_test}). The consistent performance across the two architectures underlines the portability and generality of the OpenMP-based offloading approach.

A roofline analysis on the NVIDIA platform, performed for the \textit{Classic kernel} as the production-relevant case, reveals that the its GPU implementations achieves over 80\% of the platform theoretical FP64 peak performance, operating in the compute-bound regime (see Section~\ref{Roofline_test}). This confirms not only the computational efficiency of the implementation but also the successful mitigation of thread divergence through a masked-based control strategy.

In a full production simulation setup, representative of typical large-scale simulation campaigns, the GPU-accelerated version of the \textit{Classic kernel} led to a net reduction of approximately 100 seconds per run. When extrapolated to thousands of simulations, this results in a saving of over 160000 Standard-h, representing a substantial gain in computational efficiency and resource optimization (see Section~\ref{Production_run_results}). At the same time, the scientific integrity of the \textit{Classic kernel} was extensively validated: the distribution of collapsed particles as a function of redshift showed excellent agreement with the CPU reference, with residuals typically within ±1\%, and minor systematic deviations confined to specific redshift ranges (see Section~\ref{Production_run_results}). Furthermore, the resulting numerical HMF derived from the GPU run matched the CPU-derived counterpart well below the 1\% level across the entire mass range (see Appendix~\ref{HMF_comparison}). 

Beyond the computational savings, such reductions also imply a significant potential impact in terms of energy consumption; however, a quantitative assessment of this aspect requires a dedicated analysis presented in~\cite{LACOPO2026101060} companion work.

In conclusion, this study demonstrates the feasibility and effectiveness of accelerating scientific simulation kernels on modern GPU architectures using OpenMP directives. The combination of portability, computational efficiency, and scientific accuracy affirms OpenMP as a viable long-term strategy for modernizing legacy cosmological codes and adapting them to heterogeneous HPC environments.

\section*{Acknowledgment}
This work has been supported by the Spoke-1, Spoke-2 and Spoke-3 "FutureHPC \& BigData” of the ICSC – Centro Nazionale di Ricerca in High Performance Computing, Big Data and Quantum Computing – and hosting entity, funded by European Union – NextGenerationEU. Supported by Italian Research Center on High Performance Computing Big Data and Quantum Computing (ICSC), project funded by European Union - NextGenerationEU - and National Recovery and Resilience Plan (NRRP) - Mission 4 Component 2 within the activities of Spoke 2 (Fundamental Research and Space Economy), (CN 00000013 - CUP C53C22000350006). We acknowledge the Pawsey Supercomputing Research Centre for the availability of high performance computing resources, support and collaborations. The code development, computational runs, and post-processing has been done on the Leonardo Booster platform, made available through the CINECA Italian national HPC facility. This work has been supported by the ISCRA-C project, SCGPCT (HP10CJLAHQ), and the ISCRA-B project, "Simulating thousands of Euclid skies" (EuMocks) (1498509). This work has also been supported by the National Recovery and Resilience Plan (NRRP), Mission 4, Component 2, Investment 1.1, Call for tender No. 1409 published on 14.9.2022 by the Italian Ministry of University and Research (MUR), funded by the European Union – NextGenerationEU– Project Title ”Space-based cosmology with Euclid: the role of High-Performance Computing” – CUP J53D23019100001 - Grant Assignment Decree No. 962 adopted on 30/06/2023 by the Italian Ministry of Ministry of University and Research (MUR).
\bibliographystyle{elsarticle-harv} 
\bibliography{biblio}

\appendix
\section{Halo Mass Function comparison: GPU vs CPU in a production run scenario}
\label{HMF_comparison}
\begin{figure*}[h!]
    \centering
    \includegraphics[width=0.65\textwidth]{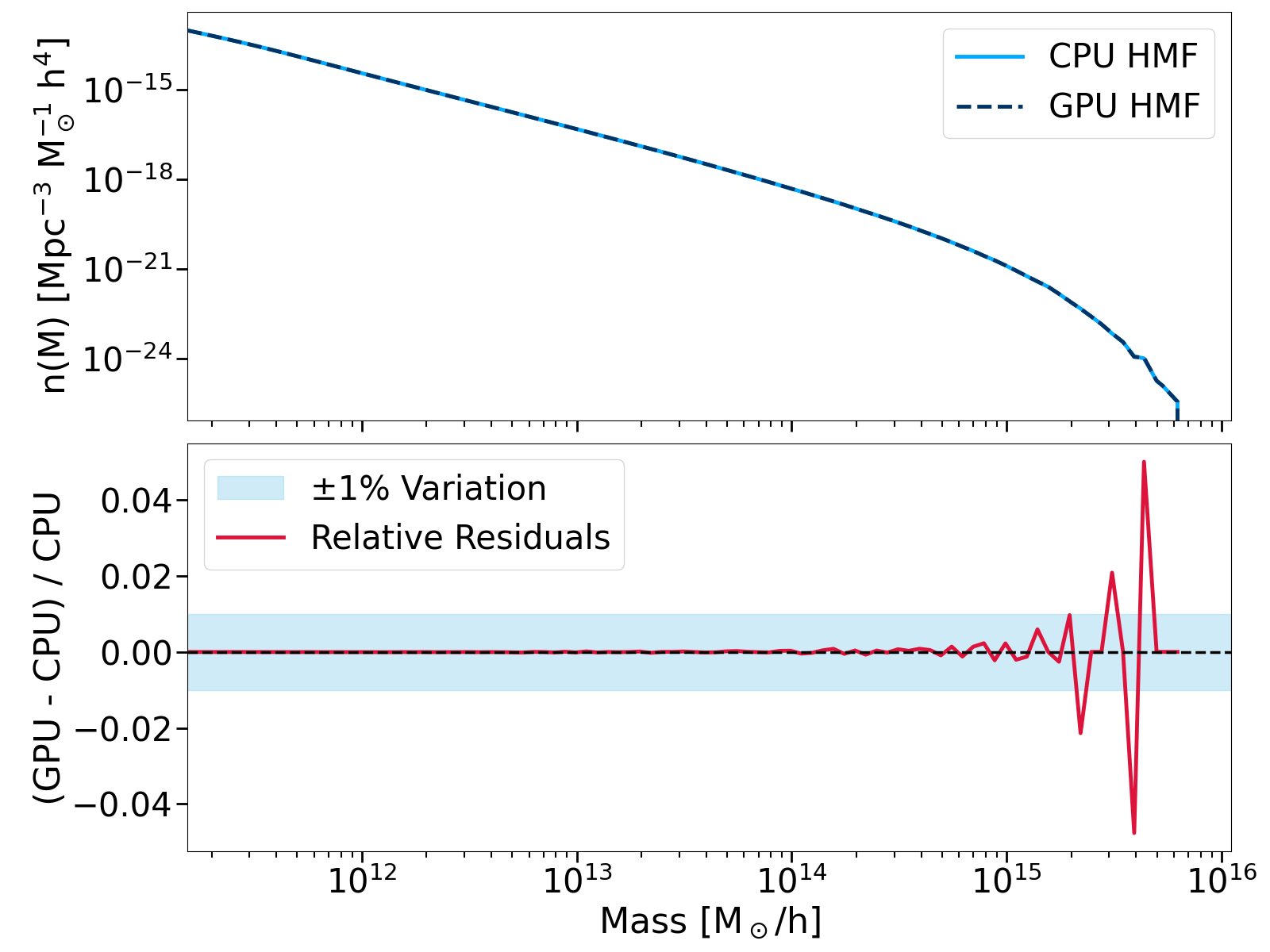}
    \caption{Comparison of the HMF obtained from the GPU-accelerated and CPU-based implementations of the {\pinocchio} \emph{Classic kernel} in a full production run. The top panel shows the HMFs from both versions, while the bottom panel displays the relative difference between them.
    }
    \label{HMF_plot}
\end{figure*}
The Halo Mass Function (HMF) is a key statistical measure in cosmology that describes the number density of dark matter haloes as a function of their mass and redshift~\citep{Press:1973iz, Sheth:1999mn,Euclid:2022dbc}. It provides a fundamental link between theoretical models of structure formation and observations, and is commonly used to validate the accuracy of cosmological simulations.

To further validate the scientific consistency of our GPU-accelerated implementation of the \textit{Classic kernel}, we compare the HMFs obtained from both the CPU and GPU versions of \pinocchio\ in a full production run scenario as described in Section~\ref{Production_run_results}. Figure~\ref{HMF_plot} presents the results. 

The agreement between the two implementations is remarkably good, with relative differences well below the 1\% level across the entire mass range. This confirms the numerical consistency of the GPU porting in terms of its impact on \pinocchio\ scientific final output. At the high-mass end, slightly larger fluctuations are observed, which are expected due to the inherently lower number statistics in that regime. These deviations do not indicate a systematic bias and are fully consistent with stochastic sampling noise. The overall agreement reinforces the reliability of the GPU-native implementation not only at the interpolation level, but also in preserving the physical accuracy of the full pipeline when applied to large-scale cosmological workflows.

\end{document}